# Literature Based Discovery (LBD): Towards Hypothesis Generation and Knowledge Discovery in Biomedical Text Mining

**Balu Bhasuran[1], Gurusamy Murugesan[2], Jeyakumar Natarajan[1,2]***

[1]DRDO-BU Center for Life Sciences, Bharathiar University Campus, Coimbatore, TamilNadu,India

[2]Data mining and Text mining Laboratory, Department of Bioinformatics, Bharathiar University, Coimbatore, TamilNadu,India

***Corresponding Author:** n.jeyakumar@yahoo.co.in

## Abstract

Biomedical knowledge is growing in an astounding pace with a majority of this knowledge is represented as scientific publications. Text mining tools and methods represents automatic approaches for extracting hidden patterns and trends from this semi structured and unstructured data. In Biomedical Text mining, Literature Based Discovery (LBD) is the process of automatically discovering novel associations between medical terms otherwise mentioned in disjoint literature sets. LBD approaches proven to be successfully reducing the discovery time of potential associations that are hidden in the vast amount of scientific literature. The process focuses on creating concept profiles for medical terms such as a disease or symptom and connecting it with a drug and treatment based on the statistical significance of the shared profiles. This knowledge discovery approach introduced in 1989 still remains as a core task in text mining. Currently the ABC principle based two approaches namely open discovery and closed discovery are mostly explored in LBD process. This review starts with general introduction about text mining followed by biomedical text mining and introduces various literature resources such as MEDLINE, UMLS, MESH, and SemMedDB. This is followed by brief introduction of the core ABC principle and its associated two approaches open discovery and closed discovery in LBD process. This review also discusses the deep learning applications in LBD by reviewing the role of transformer models and neural networks based LBD models and its future aspects. Finally, reviews the key biomedical discoveries generated through LBD approaches in biomedicine and conclude with the current limitations and future directions of LBD.

**Summary**

- Literature Based Discovery (LBD) approaches proven to be successfully reducing the discovery time of potential associations that are hidden in the vast amount of scientific literature.
- This review focuses various tools and techniques introduced in LBD using literature resources such as MEDLINE, UMLS, MESH, and SemMedDB.
- This review also discusses the deep learning applications in LBD by reviewing the role of transformer models and neural networks based LBD models and its future aspects.
- Finally, reviews the key biomedical discoveries generated through LBD approaches in biomedicine and conclude with the current limitations and future directions of LBD.

## Introduction

The major outcomes and insights of scientific research and clinical study end up in the form of publication or clinical record which as an unstructured text [1]. Due to advancements in biomedical research, the growth of published literature is getting tremendous momentum in recent years [2]. The scientists and clinical researchers facing a big challenge to stay current and to extract hidden information from this sheer quantity of millions of published biomedical literatures [3]. This large trove of unstructured data poses various problems such as data collection, management, exploration, deriving new information or discovery of knowledge, the potential one-stop automated solution to this problem is biomedical text mining [4].

According to Marti A. Hearst TM can be defined as "the discovery by computer of new, previously unknown information, by automatically extracting information from different written resources" [5]. TM is the process of generating high-quality information in the form of a novel, relevant and interesting patterns, trends, facts or hypotheses by shifting through a large volume of unstructured data [6-10]. The process of TM pipeline consists of, Information Retrieval (IR), Information Extraction (IE), and Knowledge Discovery and Hypothesis Generation [11-15]. In the context of TM, IR is the process of finding relevant natural language text from a set of literature-based databases. Normally IR is performed as a query-based or document-based search for retrieving abstract or full-text from digital libraries or databases [16-18]. IE can be defined as the automatic process of extracting structured information from semi-structured and/or unstructured machine-readable text [19]. The sole purpose of automated TM is the discovery of new knowledge, generation of new ideas or hypothesis from literature [20].

Biomedical Text Mining is concerned with the extraction of information regarding biologic entities and its relationships, such as genes and proteins, diseases, drugs, cell type, miRNA, phenotypes, or even more broadly biologic events and pathways from the scientific text [21-25].

Furthermore, the extracted information has been used for hypothesis generation, knowledge discovery, annotation of specialized databases, tools and manual curation of biological databases such as infer novel relationships: fish oil and Raynaud disease, magnesium deficiency and migraine, creation of databases CTD, OMIM, DisGeNET, STRING, building sophisticated web servers PubTator, mirCancer, PolySearch, DISEASES, PKDE4J, and formation of discovery platforms such as BEST, DigSee, Beegle, and Implicitome [26-34]. Thus, BTM has become an integral part of many resources serving a wide audience of researchers and scientists [35-42]. Figure 1 depicts schematic flowchart of an approach for identifying, screening, and including relevant studies for this review. In general, the Biomedical Text Mining pipeline consists of the fundamental three steps of text mining namely Information Retrieval (IR), Information Extraction (IE) and Knowledge Discovery from Text (KDT). The schematic architecture of a typical biomedical text mining pipeline is depicted in figure 2.

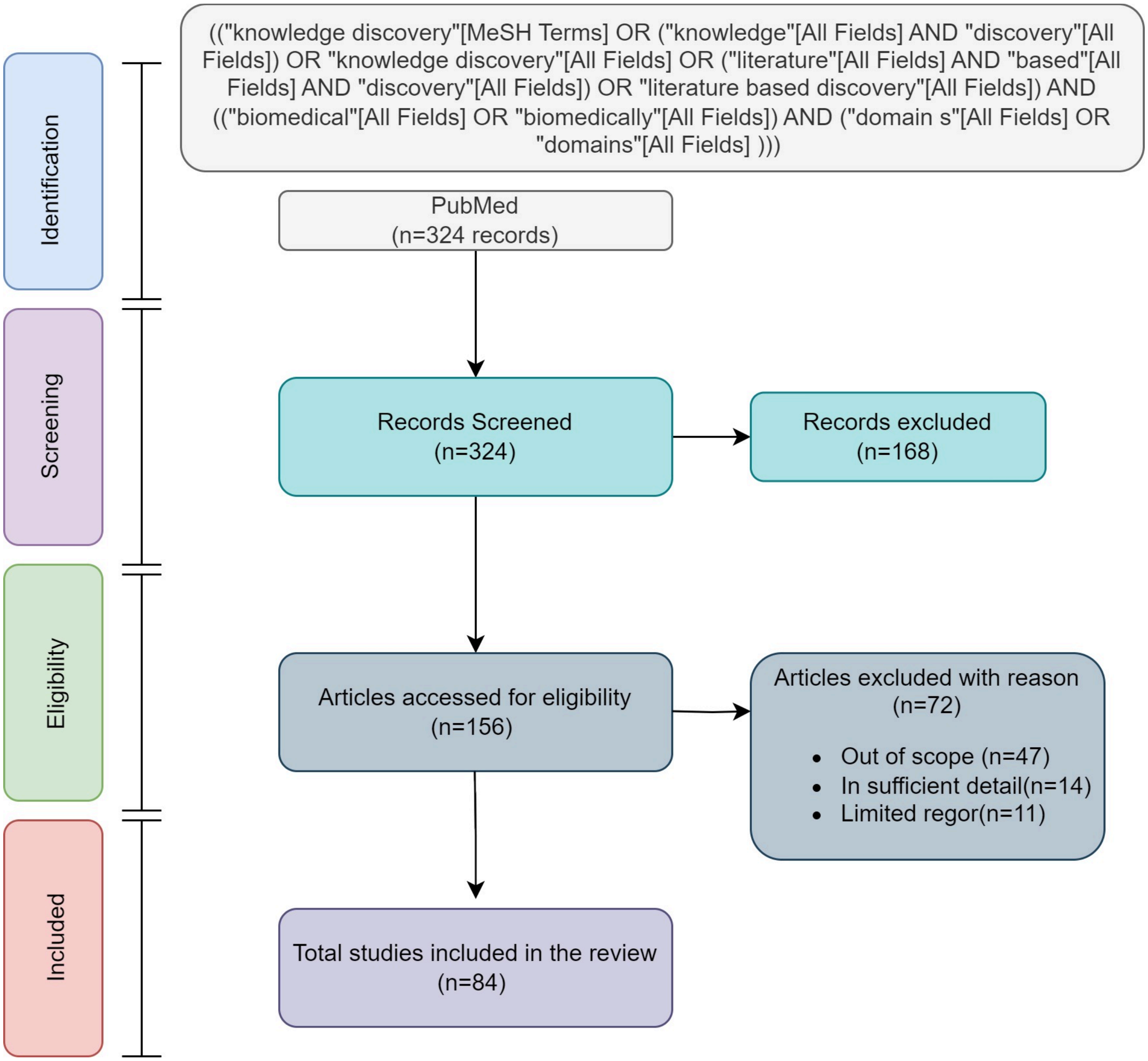

**Figure 1:** Schematic flowchart of an approach for identifying, screening, and including relevant studies

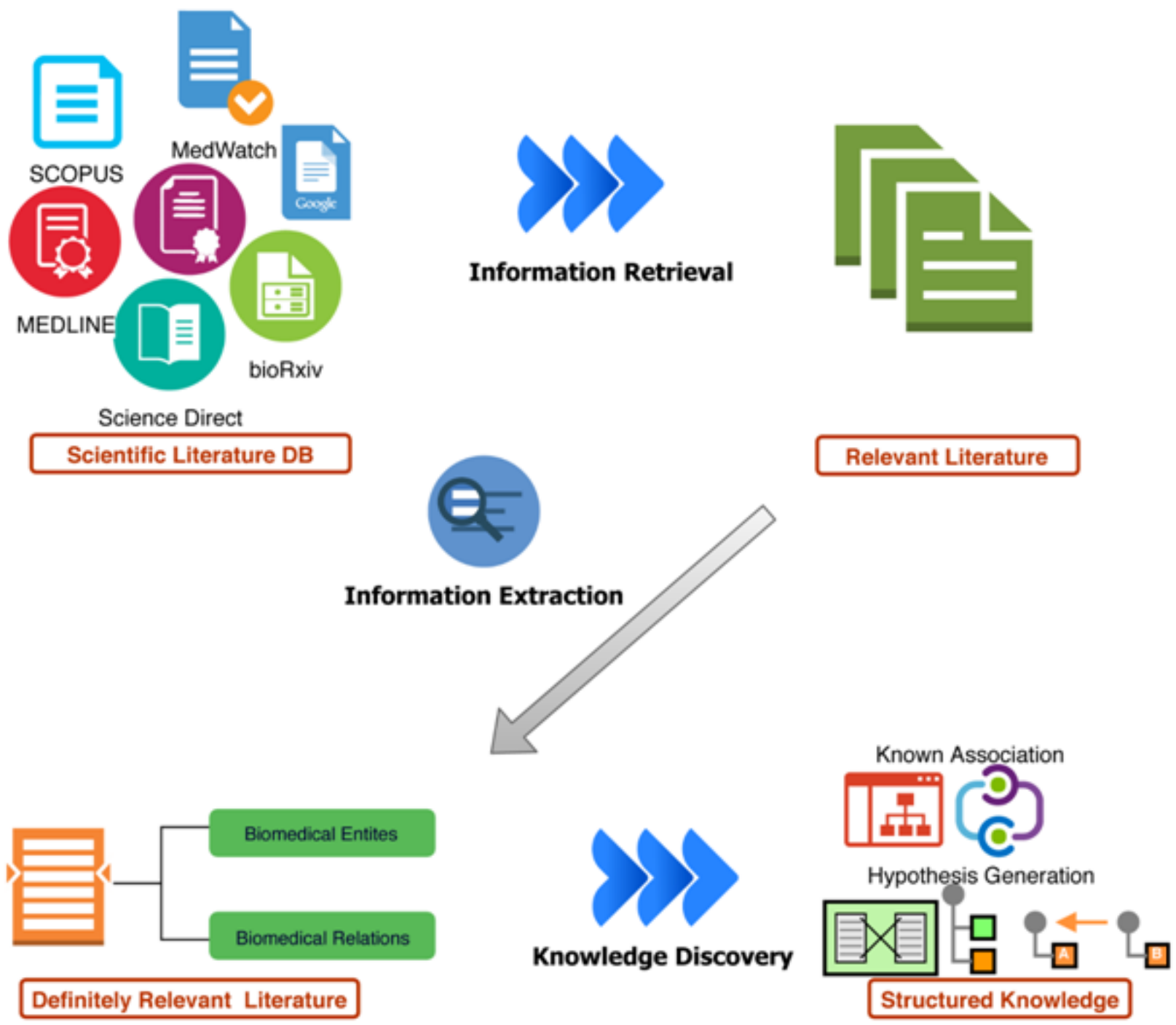

**Figure 2:** The schematic architecture of a typical biomedical text mining pipeline for hypothesis generation and knowledge discovery.

The rapid growth of scientific literature in recent years, especially in the life science domain, causes the problem of missing a large number of important connections between biomedical entities. In order to discover this unnoticed and hidden information, automated biomedical text mining procedures are required [43]. The major goal of this text mining sub-field is to formulate a hypothesis with high true probability by navigating through and connecting a large number of biomedical concepts of disjoint literate sets, which falls under Knowledge Discovery in Text (KDT) approach [44]. The discovery platforms and literature wide analysis studies (LWAS) focusing on alleviating these problems are collectively known as Literature Based Discovery (LBD) in text mining [45].

In BTM, Knowledge Discovery process is performed as a novel connection between medical terms or biological network analysis, prioritization methods. disease-specific case studies, drug search for cancers by integrating pathways and molecules, gene prioritization, global disease network generation are some major examples of these kinds in BTM [46]. Knowledge Discovery is sometimes referred to as hypothesis generation. Hypothesis generation is the process of generating unknown facts by utilizing information discovered with the use of IR and IE. Generating hypotheses in the biomedical field is a significant task to infer unknown biomedical facts that can be used to guide the design of experiments in the future or explain existing experimental results [47]. This, in turn, helps to determine new drug targets or new novel interactions between biomedical concepts which is not proved before [48].

One of the well-known approaches in this task is proposed by D.R. Swanson using the ABC principle to link disjoint literature sets for biomedical knowledge discovery [49]. The ABC principle states that if concept A and concept B was associated directly in one set of literature, while concept B and concept C was in direct relation to an independent disjoint set, then the union of these literature sets allows a new possible inference relation between concepts A and C linking via the concept B. Example 1 illustrates the ABC principle by linking magnesium deficiency to migraine [50].

Example 1: D R Swanson [49] identified that the medical concepts stress, calcium channel blocker, spreading cortical depression (SCD), and platelet aggregability connects the magnesium deficiency to migraine. He used literature pieces of evidence such as stress is associated with migraine and stress can lead to loss of magnesium, calcium channel blocker prevents migraine and magnesium is a natural calcium channel blocker, etc. Later the discovery is validated experimentally [51].

Kastrin and Hristovski did the first inclusive scientometric overview of the LBD study covering 35 years (1986–2020) using 409 documents from six bibliographic databases [52]. The overview found Rindflesch TC, Kostoff RN, Hristovski D, Smalheiser NR, and Swanson DR as the top five authors in LBD based on number of publications. The study also generated a Co-authorship network, Document co-citation network in this domain and top journals publish the studies in LBD domain. We recommend this study for a better understanding of LBD in biomedicine, the origin and evolving [52].

## Literature Source

## MEDLINE

One of the well-known sources of biomedical literature articles is MEDLINE and its widely used for worthwhile to mention that is the most popular source of data for literature collection in biomedical domain. As there is limited availability of full text articles, most researchers use MEDLINE title, abstract and indexing terms (MeSH terms) as a surrogate for full text articles. Also, an advantage of using MEDLINE is that it is openly accessible and can be searched using a powerful search engine developed by National Library of Medicine (NLM), viz., PubMed. The articles in MEDLINE contain several attributes, a few of the widely used are listed here: PMID (a single element to uniquely identify articles), Article Title (the title of each article), Abstract (abstract text of each article), PubDate (it contains the full date on which the article was published), MeshHeadingList (it contains the MeSH terms assigned for each article).

## Auxiliary knowledge sources

**Unified Medical Language System**

Unified Medical Language System (UMLS) is one of the major biomedical knowledge bases used in Biomedical and health informatics domain. UMLS enable semantic understanding and interoperability among different software applications and systems by combining widely used dictionaries in the biomedical field. UMLS contains three knowledge sources namely i) Metathesaurus ii) Semantic Network iii) SPECIALIST Lexicon

i) ***Metathesaurus:*** It is the main component of the UMLS and it is organized by combining various biological concepts (such as Gene, Protein, Disease names etc.). UMLS utilizes metathesaurus to connect the alternative names of the same concepts from various sources of dictionaries. Metathesaurus not only links the same concepts from various sources but also used to identify relationships among different concepts.

ii) ***Semantic network:*** The biological concepts described in the UMLS Metathesaurus are grouped into subject categories called semantic types. For example: The concept breast cancer belongs to semantic type[ "Disease or Syndrome"] and magnesium categorized as ["chemical"]. UMLS also contains the relation between these semantic types called "semantic relations".

iii) ***SPECIALIST lexicon:*** The SPECIALIST Lexicon contains the information (word usage) used by the Natural Language Processing (NLP) processing. Each entry in this lexicon includes the morphological, syntactic and orthographic information for each word or term.

**Medical Subject Headings(MESH)**

Medical Subject Headings (MESH) was introduced by NLM for indexing and retrieval of PubMed articles by using Supplemental Concept Records (SCR) and Mesh terms for each article. Generally, Mesh terms provides abstract or summarized biological concepts used in the paper.

MESH terms are classified into three sub-types i) Descriptors: it denotes the main concepts of the article described for example if the article describe the role of magnesium deficiency in the role of neurological disorders the is categorized and indexed in “Magnesium Deficiency” and “Neurological disorders”. Descriptors are standalone terms compared to other terms. ii) Qualifiers: mainly useful if it is used in conjunction with descriptors. iii) Supplementary Concept Records (SCR): SCR index named entities associated with the article such as gene, disease name, chemical etc. Apart from the above three sub-types MESH also contains a code called mesh tree code which is arranged in a hierarchical manner. Thus, MESH concept provides effective way of searching for the articles of specific biomedical subjects.

**SemMedDB**

Semantic relations are important for text mining tasks such as knowledge discovery and hypothesis generation. SemMedDB [53] is the repository of semantic relation extracted from PubMed articles title and abstracts using the rule based system called SemRep. SemMedDB Contains the predictions of (subject-predicate-object) triples from the PubMed articles. SemMedDB uses UMLS Metathesaurus for concept extraction and for relation extraction, it uses Semantic Network concept. The Semantic MEDLINE Database (SemMedDB) indexes semantic predications triples (subject-predicate-object ) extracted by the a semantic interpreter SemRep from PubMed citations. The current version of SemMedDB provides approximately 96.3 million predications from SemRep using 29.1 million citations from MEDLINE database. This PubMed scale MySQL database provides information about the PubMed citation, One-to-many relationships of the concept with UMLS metathesaurus information, links between predictions and between a prediction and a sentence.

Literature sources and NLP tools in text mining for LBD is given in table 1 and table 2 provides curated knowledge sources for LBD in biomedicine below.

**Table 1:** Literature sources and NLP tools in text mining for LBD

| Type | Name | Web-Link | Type | Current Status |
|---|---|---|---|---|
| Literature Sources | MEDLINE | https://www.nlm.nih.gov/medline/index.html | Online DB | Working |
| | Scopus | https://www.elsevier.com/en-in/solutions/scopus | Online DB | Working |
| | Science Direct | https://www.sciencedirect.com/ | Online DB | Working |
| | Europe PMC | https://europepmc.org/ | Online DB | Working |
| | bioRxiv | https://www.biorxiv.org/ | Online DB | Working |
| NLP Tools | SemRep | https://semrep.nlm.nih.gov/ | Standalone/ Downloadable | Working |
| | MetaMap | https://metamap.nlm.nih.gov/ | Standalone/ Downloadable | Working |
| | cTAKES | https://ctakes.apache.org/ | Standalone/ Downloadable | Working |

**Table 2:** Curated Knowledge sources for LBD in Biomedicine

| Curated Knowledge Source | Web-Link | Type | Current Status |
|---|---|---|---|
| **OMIM** (Online Mendelian Inheritance in Man) (*Hamosh et al., 2005*) | http://www.ncbi.nlm.nih.gov/omim | Online DB | Working |
| **CTD** (Comparative Toxicogenomics Database) (*Davis et al., 2018*) | http://ctd.mdibl.org | Online DB | Working |
| **STRING** (*Szklarczyk et al., 2017*) | http://string-db.org/ | Online DB | Working |
| **DisGeNET** (*Pinero et al., 2017*) | http://www.disgenet.org | Online DB | Working |
| **PharmGKB** (Pharmacogenomics Knowledgebase) (*Thorn et al., 2013*) | http://www.pharmgkb.org | Online DB | Working |
| **UniProt** (*The UniProt Consortium, 2018*) | http://www.uniprot.org/ | Online DB | Working |
| **MEDIC** (merged disease vocabulary) (*Davis et al., 2012*) | http://ctd.mdibl.org/voc.go?type=disease | Online DB | Working |
| **DO** (Disease Ontology) (*Kibbe et al., 2015*) | http://www.disease-ontology.org | Online Tool | Working |
| **UMLS** (Unified Medical Language System) (*Olivier Bodenreider, 2004*) | http://umlsks.nlm.nih.gov | Standalone/ Downloadable | Working |
| **MeSH** (Medical Subject Headings) (*CE Lipscomb,* | https://www.nlm.nih.gov/mesh/ | Online Tool/ Standalone | Working |
| *2000*) | | | |
| **SNOMED CT** (Systematized Nomenclature of Medicine-Clinical Terms) (*Elkin et al., 2006*) | http://www.snomed.org/ | Online DB | Working |

## Literature Based Discovery Methods

This section briefly discusses the various terms and concepts explored in LBD such as ABC principle, Concept profile, open and closed discovery process followed by key applications and discoveries in biomedical text mining domain

### ABC Principle

Using biomedical text mining integrated with prior knowledge (already known biomedical associations), intuition and experience scientific discoveries are made in biomedicine. Text Mining plays an important role by assisting this process using automatically discovering interesting novel hypothesis. In this regard text mining researchers adapted the widely explored ABC principle for hypothesis generation and knowledge discovery [50]. The ABC principle can be stated as, if concept A and concept B was associated directly in one set of literature, while concept B and concept C was in direct relation to an independent disjoint set, then the union of these literature sets allows a new possible inference relation between concepts and linking via the concept B [54]. This approach enables to establish a new link between concept A to Concept C via shared concept B. The concepts can be any relevant biological entity such as gene, protein, drug, disease, cell line, miRNA, antibody, peptide or common medical terms from repositories like MeSH [55].

The two core aspects of the ABC principle that is explored widely in the LBD studies are the identification of shared(related) interesting(promising) concepts and exploring those relationships based on certain particulars (biologically relevant). For example, that starting point of the search can be a disease X [56]. The first step of the approach will be identifying related medical concepts to X such as a particular biomarker gene Y1 or drug Y2. In the second step, the relationship is explored in a such a way that is there a mutation in Y1 gene causes disease X, is drug Y2 likely to treat disease X or aggravates X [57-59]. The aim of the approach is to derive some kind of novelty in the relationship and discovering more Y concepts (chemical, miRNA or medically relevant reactions) there by creating more new possible hypothesis between X and Y [60-62]. According to the way the concepts are searched, the approaches are classified as open discovery and closed discovery. An open discovery process aims for a hypothesis generation by navigating through connected concepts at different levels [63-67]. A closed discovery process starts with known concepts at both ends A and C respectively. In this process the approach searches for B terms that can support the claim that A-C association is a relevant one [68-72]. A schematic representation of open and closed discovery approaches in literature based discovery is depicted in figure 3.

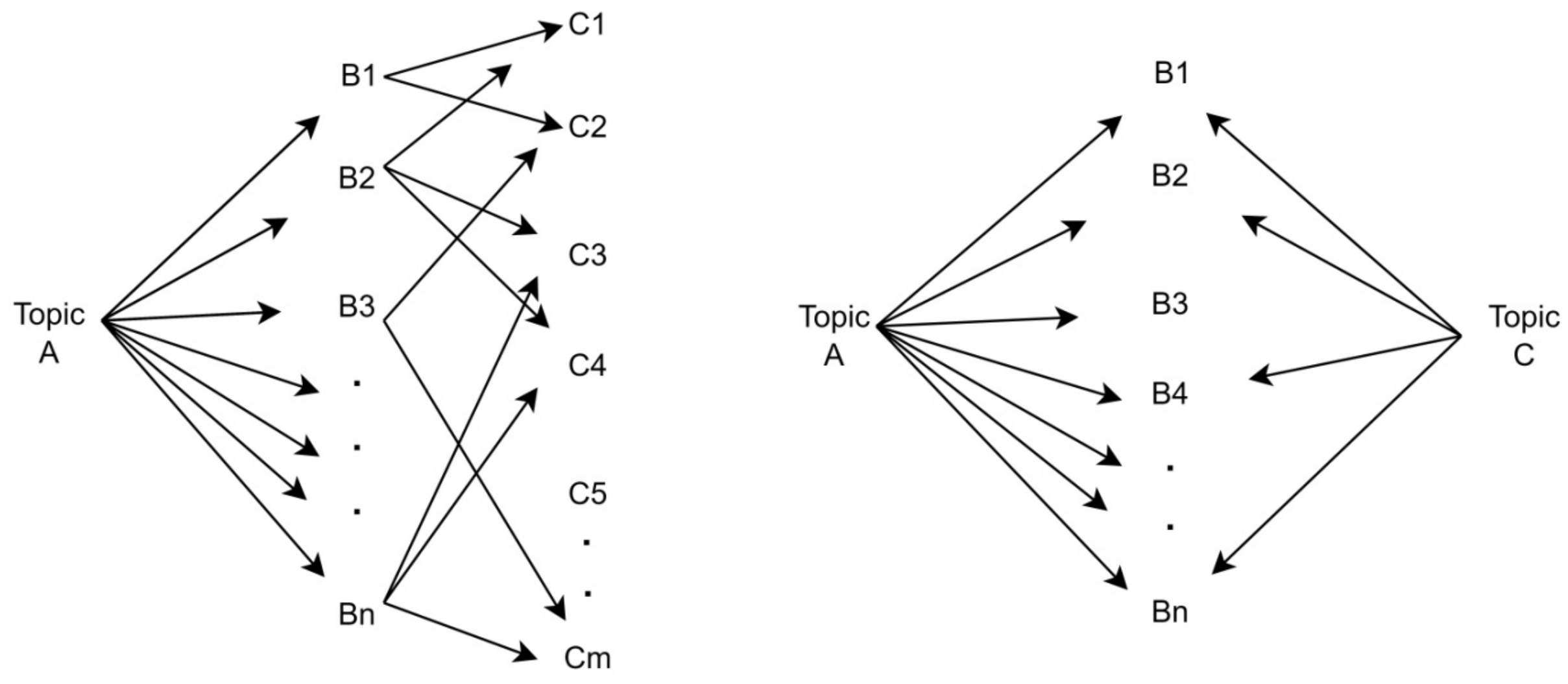

**Figure 3:** Open and Closed Discovery Approaches in Literature Based Discovery defined by Weeber at al. [81]

**Concept Profile**

Concept Profile of a biological entity represents a set of terms that are related to the entity either through curated known association or through a co-occurrence mention in a biologically relevant context [73]. Consider a topic such as Alzheimer's disease, which is an irreversible, progressive brain disorder. The profile for this topic distilled from a suitable text collection could identify, for example, terms representing the genes, proteins, symptoms, drug treatments, other diseases, and population groups associated with the disease, i.e., “statistically related” to it. In majority of the cases biologically co-occurrence implies semantic association [74-77]. One way to create a concept profile is to apply MeSH metadata on MEDLINE databases using dictionary matching or automatic concept identifiers such as MetaMap or cTAKES. A concept profile can be represented as,

$$\mathbf{Profile(Ti)= \{w_{i,1}.m_1, w_{i,2}.m_2, \ldots, w_{i,n}.m_n\}} \qquad \mathbf{(1)}$$

j=(1,2,...n)

where $m_j$ represents a MeSH term, $w_{i,j}$ its weight, and there are totally n terms in the MeSH vocabulary.

**Open Discovery**

An open discovery process aims for a hypothesis generation by navigating through connected concepts at different levels. Initially, there is only the starting concept that can be a scientific problem or research question and the end of the discovery is not defined. For example in Swanson initial study was to find a new treatment for Raynaud's disease [49-50]. Using discovery approach uses disease C as the initial concept searches for interesting clues (B), typically treating drugs, molecular pathways, or physiological processes that play a role in the disease under scrutiny. Next, the approach finds A-terms, typically substances/drugs or pathways, that act on the selected Bs. Major challenge of open discovery support tools is to contain the vast amount of possibilities identified in the initial search. Finally, a hypothesis can be formulated like the substance A can be used for the treatment of disease C. Since search space is expanded in to multiple levels due to connected concepts, a better understanding of the problem with domain knowledge is necessary for open discovery [78-80].

**Closed Discovery**

A closed discovery process starts with known concepts at both ends A and C respectively. In this process the approach searches for B terms that can support the claim that A-C association is a relevant one. For the sample example as discussed above the closed discovery approach starts from both disease C and substances/drugs A, the approach searchers for common intermediate B-terms. The more pathways or physiological processes between A and C the search results, the more likely this hypothesis is a valid one [81-83]. Due to the simplicity and better search paradigm most of the LBD approaches are focused on closed discovery. Since both the concepts are already known, the approaches simply search for B-terms between them. Various association hypothesis such as gene-disease and drug-disease is generated using closed approach [84-89]. Novel potentially relevant interesting and spurious biological link identification using Open and Closed discovery approaches is depicted in figure 4.

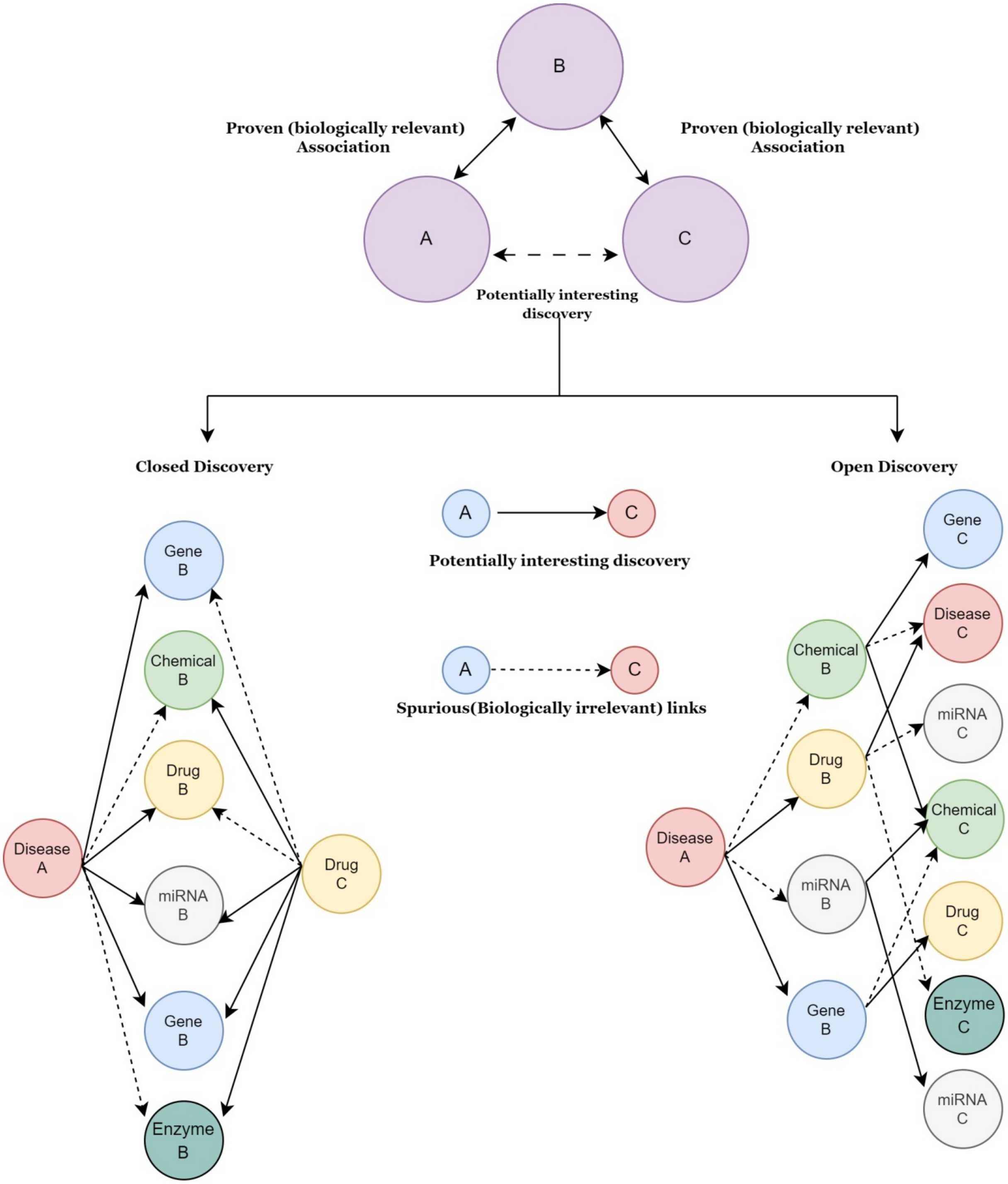

**Figure 4:** Novel potentially relevant interesting and spurious biological link identification using Open and Closed discovery approaches. Through closed discovery disease A and Drug C is connected through gene B, miRNA B where as other discoveries are not relevant. Through open discovery Disease A to Drug C is connected via Gene B, Disease A to Disease C is connected via Drug B, other connections are not relevant.

## Applications in Biomedical Text Mining

Since the amount of available biomedical research literature is overwhelming, the scientist faces a greater challenge to navigate through all the relevant articles of a specific disease, gene, chemical or miRNA of interest to formulate a research hypothesis or derive a future connection. One of the initial breakthrough studies focusing LWAS was performed by Swanson DR in which he created a hidden relationship model by searching through disjoint literature sets and formulated multiple novel hypotheses such as biological connection between magnesium and migraine, fish oil and Raynaud's syndrome, and somatomedin C and arginine [49-50,53]. All of these connections were never predicted or reported earlier and proved later which opened the door for a new era in biomedical text mining. Following this root and adapting the famous ABC principle Hristovski et al, released BITOLA, MEDLINE database based meaningful relation generator using user given MeSH term as pivot concepts [65]. The web server expects the user to give a meaningful concept and incorporates external knowledge sources such as a chromosomal location for performance improvement [65]. Another well-known text mining-based knowledge discovery system developed by Smalheiser et al named Arrowsmith, using B-term phrases and title words connecting the articles with a two-node approach-based searching[90]. Another major real-time discovery tool FACTA+ created by Tsuruoka et al, based on concept co-occurrence at the abstract level integrating hidden association generation, biomolecular events and network visualization[91]. Fleuren et al developed CoPub 5.0, an integrative framework with co-occurrence and keyword-based searching, ABC principle based hidden connection, and Cytoscape software-based network construction [18]. CoPub 5.0 has three search modes namely term search(to retrieve abstract and keyword relation extraction for a particular term), pair search(Analyze the new relation or known relation) and set of terms(relation between multiple terms) to answer biological questions . Figure 5 shows various approaches and examples in LBD systems such as co-occurrence bases, semantic relation based, graph based and hybrid approaches. Table 3 represents details of knowledge discovery tools using biomedical text mining sources. Table 4 shows the major literature based discoveries in biomedicine using ABC principle.

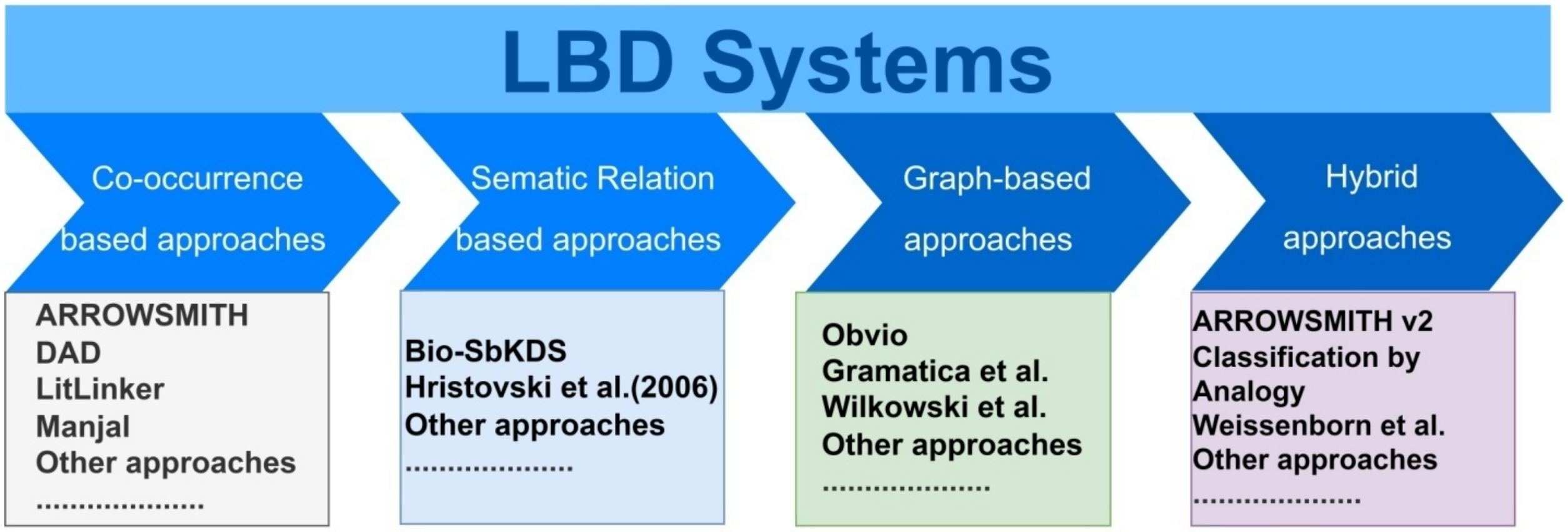

**Figure 5:** Various approaches and examples in LBD systems

**Table 3:** Detailed representation of knowledge discovery tools using biomedical text mining

| **Discovery Tool** | **Web-Link** | **System Description** | **Type** | **Current Status** |
|---|---|---|---|---|
| DigSee (*Kim et al., 2017*) | http://gcancer.org/digsee | Direct(explicit) gene-disease associations from genes involved in the bio-molecular events with sentence scoring | Online Tool | Inoperative |
| LION LBD (*Pyysalo et al., 2018*) | http://lbd.lionproject.net | Implicit and Explicit associations generation using mapped ontology and concept graph with a special emphasis on Cancer | Online Tool | Working |
| The Implicitome (*Hettne et al., 2016*) | http://knowledge.bio | Indirect(implicit) gene-disease associations using concept profiles using ABC principle and association score | Online Tool | Inoperative |
| Textpresso Central (*Muller et al., 2018*) | http://www.textpresso.org/tpc | In depth search and annotation tool with customization and integration option | Online Tool | Inoperative |
| Beegle | http://beegle.esat.kuleuven.be/ | Implicit and | Onlin | Inoperativ |
| (*ElShal et al., 2015*) | | explicit associations identified through co-occurrence and concept profile, integrated with a prioritization tool | e Tool | e |
| CoPub 5.0 (*Fleuren et al., 2011*) | http://www.copub.org | Integrative framework with co-occurrence and keyword-based searching, ABC principle based hidden connection | Onlin e Tool | Inoperativ e |
| GS2D (*Miguel A. Andrade-Navarro and Jean Fred Fontaine,* | http://cbdm.uni-mainz.de/geneset2diseases | Direct(explicit) gene-disease associations with co-occurrence statistics and disease enrichment | Onlin e Tool | Available |
| *2016*) | | analysis | | |
| FACTA+ (*Tsuruoka et al., 2011*) | http://refine1-nactem.mc.man.ac.uk/facta/ | Concept co-occurrence at abstract level integrating hidden association generation, bio molecular events and network visualization | Online Tool | Inoperative |
| Anni 2.0 (*Jelier et al., 2008*) | http://www.biosemantics.org/anni | Implicit and Explicit associations with co-occurrence and ontologies | Online Tool | Inoperative |
| Arrowsmith (*Smalheiser et al., 2009*) | http://arrowsmith.psych.uic.edu | B-term phrases and title words connecting the articles with a two-node approach-based searching | Online Tool | Working |
| FACTA (*Tsuruoka | http://www.nactem.ac.uk/software/facta/ | Direct(explicit) associations with | Online Tool | Working |
| *et al., 2008*) | | co-occurrence statistics and point wise mutual information | | |
| PolySearch (*Cheng et al., 2008*) | http://wishart.biology.ualberta.ca/polysearch | Large number of dictionaries and bag-of-words for direct(explicit) associations | Online Tool | Working |
| DISEASES (*Pletscher-Frankild et al., 2015*) | http://diseases.jensenlab.org/ | Direct(explicit) associations integrated with cancer mutation data and manually curated databases | Online Tool | Working |
| PolySearch 2 (*Liu et al., 2015*) | http://polysearch.ca | Update of PolySearch with tightness measure based on word position | Online Tool | Working |
| Anni (*Jelier et al., 2007*) | http://www.biosemantics.org/Anni | Concept profile weighting using | Online Tool | Inoperative |
| | | likelihood ration | | |
| BITOLA (*Hristovski et al., 2005*) | http://www.mf.uni-lj.si/bitola/ | User given MeSH term as pivot concepts with external knowledge sources such as chromosomlal location | Online Tool | Working |
| iTextMine (*Ren et al., 2018*) | http://research.bioinformatics.udel.edu/itextmine | Automated work flow with parallel processing for explicit associations | Online Tool | Working |
| DEXTER (*Gupta et al.,2018*) | http://biotm.cis.udel.edu/DEXTER | Disease expressions extraction with co-occurrence and argument filtering | Online Tool | Working |
| MELODI (*Elsworth et al., 2018*) | www.melodi.biocompute.org.uk | Graph based database for mechanistic pathways | Online DB | Working |
| | | identification | | |

Recent work by Tropmann-Frick and Schreier discussed the various drug repurposing approached for Covid-19 using LBD shows the potential and immediate applications of the field [91]. They used three LBD systems Arrowsmith, BITOLA, and SemBT for the searching of repurposable drugs for Covid-19 using the ABC principle. With closed discovery approach using Arrowsmith they used COVID-19 and drug 'remdesivir' as A and B concepts, for open discovery they used 'molecular mechanisms of pharmacological action' as target concept. Using BITOLA they used SARS and ' chloroquine' in closed discovery and SARS and 'lactate dehydrogenase' as source and target concepts respectively. The study generated rank frequency, rank coefficient, frequency (AB, BC), novel discovery status, confidence values etc to statistical insights to the discovery [91].

One improvement to ABC principle is discussed in Baek et al., in their study for plausible new hypothesis generation from PubMed. They discussed two aspects namely context surrounding and clinical validation [92]. The proposed solution emphasized on multiple B terms in metabolite-related hypothesis with diverse biological types. The study found that lactosylceramide and arterial stiffness is associated with the involvement of a potential pathway connecting the entities and nitric oxide, malondialdehyde and they clinically validated the generated hypothesis [92]. The same author further expanded the new ABC principle to a context-based and a context-assignment-based ABC models by using four biological context elements: cell, drug, disease, and organism. This study showed that there is a 50-70% improvement in precision for identifying association between APOE–MAPT and FUS–TARDBP comparing co-occurrence based ABC model with the context-based ABC models [93].

| **Medical Entity 1 (A)** | **Medical Entity 2 (C)** |
|---|---|
| Migraine | Magnesium |
| Raynaud Disease | Fish Oil |
| Indomethacin | Alzheimer's Disease |
| Estrogen | Alzheimer's Disease |
| Calcium-Independent Phospholipase A2 | Schizophrenia |
| Magnesium deficiency | Neurologic |
| Thalidomide | Chronic Hepatitis C |
| Testosterone | Sleep |
| Somatomedin C | Arginine |
| Chlorpromazine | Cardiac Hypertrophy |
| Diethylhexyl (DEHP) | Sepsis |
| Sleep | Depression |

**Table 4:** Important literature based discoveries in biomedicine using ABC principle

## Deep Learning LBD Models

### *Transformer Models*

Contextual word embedding and transfer learning methods provided immense momentum and new dimensions to NLP. Biomedical domain adapted this momentum very quickly through various models pre trained from weights such as BioBERT [94] or generated from scratch such as ELECTRAMed [95]. The baseline for these models is the Bidirectional Encoder Representations from Transformers (BERT) architecture a bidirectional self-attention model that uses encoder layers for two tasks, masked language modeling (MLM) and next sentence prediction (NSP). Lee et al., proposed domain specific language model BioBERT (Bidirectional Encoder Representations from Transformers for Biomedical Text Mining) by further training the weights of BERT (English Wikipedia and Books Corpus) using PubMed abstracts and PMC full-text articles [94]. The BioBERT models base and large versions shown to be outperform general BERT models and other biomedical models in three tasks namely, Named Entity Recognition (NER) in 9 biomedical datasets, Relation Extraction (RE) in 3 biomedical datasets and Question Answering (QA) in 3 biomedical datasets [94]. The wider success of BioBERT enabled the researchers to develop new transfer learning models using biomedical and clinical literature resulted in PubMedBERT [96], ClinicalBERT [97], MT-clinical BERT [98], Umlsbert [99], ELECTRAMed [95], BioMegatron [100] etc. Most of these models are pretrained with various combinations of scientific literature data PubMed and PMC, clinical data MIMIC-III and biological databases MeSH and Unified Medical Language System (UMLS) Metathesaurus.

Even though these models are trained with entire biomedical literature (29 M) knowledge and are performing well in entity recognition, linking and summarization tasks, a well designed literature based discovery task is not yet widely formulated using biomedical transfer learning models. Very recently Sybrandt et al., proposed AGATHA [101], a graph based transformer model for hypothesis generation. This deep learning based system using SciBERT tested using a temporal holdout set used a data-driven ranking criteria for generating new biomedical connections. The study constructed a large semantic graphs (over 10 billion edges) encoding sentences, entities, n grams, lemmas and terms from UMLS and MeSH and predictors from SemRep, ranked term pairs and validated the generated hypothesis using Heuristic-Based Ranking, Sub domain Recommendation, Edge2Vec Comparison and Ablation Study [101]. These new approaches shows that the neural word embeddings like BERT which considers the context of a sentence, which shows high performance in multitask prediction have immense potential in LBD. Deep learning based discovery models like AGATHA which is exploiting the association of contextual vector representations and graph neural networks are opening a new dimension for researchers in biomedicine.

### *Neural Networks*

Crichton et.al proposed four graph-based, neural network methods using Large-scale Information Network Embedding (LINE) in open and closed discovery and compared the performance with LION LBD system in the context of cancer case discoveries and a Time-slicing based approach with post cut-off publication year bases evaluation sets [102]. Baseline models were generated using 8 co-occurrence metrics namely Co-occurrence count, Document count, Jaccard Index, Symmetric conditional probability (SCP), Normalized point wise mutual information (NPMI), Chi-squared ($\chi 2$), Student's t-test (t-test) and log-likelihood ratio (LLR). The neural methods are proposed as link prediction where node embeddings created using LINE along with Jaccard Index based weighted edges [102].

For closed discovery Multi-Layer Perceptron (MLP) architecture is trained as a classifier with link between A and C has be taken into account from the grpah. The first model, Closed discovery 1 (CD-1) generated a score for every A-B or B-C link where as the second model Closed discovery 2 (CD-2) A-B-C embeddings link as the single input the trained the model to assign a prediction score for the entire association A-B-C. The authors claimed that this approach give more flexibility in terms of the length of the association (the number of B entities involved in the association). The open discovery 1 (OD-1) followed the same pipeline of CD-1 with a difference in using accumulator function for ranking based on prediction scores. The final model open discovery 1 (OD-2) used a Convolutional Neural Network (CNN) with single vector input and prediction output pipeline. This model allowed the removal of aggregator and accumulator functions and along with merging many paths from A-Bs to the same C concept. A links are given as a single dimension input to CNN and generated a A-C link prediction score. In order to keep the same window input window size the authors used elementwise summation and zero padding to fill the gap. This CNN model used a ReLU activation, max pooling and Softplus activation functions [102].

This final model also gives insights that we can use the other deep learning or transformer architectures such as BERT for LBS where a token length of 512 or fixed is expected as input. This approach also shed light to the use of graph neural networks in LBD.

To summarize the deep learning based LBD approaches, we can use transformer model like BERT for hypothesis generation by encoding sentences, entities and its ngram, to which semantic predictions and UMLS and MeSH data can be infused. It is also possible to train open and closed LBD systems using MLP and CNN based deep learning architecture by formulating the ABC problem as link prediction where entities are represented with a graph neural network and A-B-C links are given as sequence input with padding and the classifiers are generated a prediction score for A-C links.  These approaches will pave path to the future direction of LBD.

## Limitations and Future Directions of LBD

LBD approaches proven to be successfully reducing the discovery time of potential associations that are hidden in the vast amount of scientific literature. This knowledge discovery approach introduced in 1989 still remains as a core task in text mining. One of the major improvements can be done for the knowledge base creation steps. Using precise NLP tools and enrichment of multiple bioinformatics database sources researchers can generate highly curated and enriched databases through which better discoveries can be performed. With all the advancement in NLP methods and tools, LBD systems still lacks the concept linkages that are true in nature. This is very crucial for LBD systems and Biomedicine. More advancement approached from knowledge graphs networks and social media mining approaches can be adapted to rectify this issue. Another important aspect is the focus towards context and shared context and ignoring the features that are distributed between concepts. The medical concepts contributing to concept profiles are given lesser priority when the context of interest is irrelevant. Other aspects of future improvement can be, advancement of logic and reasoning approached for meaningful discoveries, reduction of manual intervention and domain knowledge expertise to generate fully automated LBD pipeline, visualization of concepts in a dynamic manner etc. Finally, very recently neural networks and transformer based for LBD is published. This will definitely open a new door towards application of various deep learning architectures adapting to the LBD approaches.

**Author contributions**

All authors have made a substantial, direct, intellectual contribution to this study.

**Conflict of interest**

The authors declare no conflict of interest.

**Acknowledgment**

This work was supported by DRDO- BU Centre for Life Sciences, Coimbatore, Tamilnadu, India. BB acknowledges the fellowship received from the grant.